\documentclass[twocolumn,showpacs,aps,prl,superscriptaddress]{revtex4}

\usepackage{graphicx}
\usepackage{dcolumn}
\usepackage{amsmath}
\usepackage{epsfig}

\input pubboard/babarsym

\RequirePackage{xspace}

\newcommand{\pvec}{{\bf p}}

\newcommand{\acp}{\ensuremath{\calA_{ch}}}

\newcommand{\calB}{\ensuremath{{\cal B}}}

\newcommand{\timesix}{\ensuremath{\times10^{-6}}}

\newcommand{\DE}{\ensuremath{\Delta E}}
\newcommand{\mb}{\ensuremath{m_{\rm ES}}}
\newcommand{\mres}{\ensuremath{m_{\rm res}}}
\newcommand{\xf}{\ensuremath{{\cal F}}}
\newcommand{\hel}{\ensuremath{{\cal H}}}
\newcommand{\thetaT}{\ensuremath{\theta_{\rm T}}}
\newcommand{\costhr}{\ensuremath{\cos\thetaT}}

\newcommand\etal{{\it et al.}}
\newcommand{\half}{\ensuremath{{1\over2}}}

\newcommand{\bma}[1]{\boldmath{$#1$}}
\newcommand{\msp}{\ensuremath{\phantom{-}}}

\newcommand{\bfig}{\begin{figure}[htbpc!]}
\newcommand{\efig}{\end{figure}}
\newcommand\bef{\begin{figure}}
\newcommand\edf{\end{figure}}
\newcommand\dbline{\noalign{\vskip 0.10truecm\hrule}\noalign{\vskip 2pt}\noalign{\hrule\vskip 0.10truecm}}
\providecommand{\tbline}{\noalign{\vskip 0.05truecm\hrule\vskip0.05truecm}}

\newcommand\beq{\begin{equation}}
\newcommand\eeq{\end{equation}}
\newcommand\bear{\begin{array}}
\newcommand\enar{\end{array}}
\newcommand\beqa{\begin{eqnarray}}
\newcommand\eeqa{\end{eqnarray}}
\newcommand\ben{\begin{enumerate}}
\newcommand\een{\end{enumerate}}

\newcommand{\UfourS}{\ensuremath{\Upsilon(4S)}}

\newcommand{\etagg}{\ensuremath{\eta_{\gaga}}}
\newcommand{\etappp}{\ensuremath{\eta_{3\pi}}}
\newcommand{\etatogg}{\ensuremath{\eta\ra\gaga}}
\newcommand{\etatoppp}{\ensuremath{\eta\ra\pi^+\pi^-\pi^0}}

   \newcommand{\KstpKppiz}{\ensuremath{\Kstarp_{K^+\pi^0}}}
   
   \newcommand{\KstpKspip}{\ensuremath{\Kstarp_{\KS\pi^+}}}
   
   \newcommand{\KstzKppim}{\ensuremath{\Kstarz_{K^+\pi^-}}}

\newcommand{\kzs}{\ensuremath{\KS}}

\newcommand{\fetaKst}{\ensuremath{\eta K^{*}}}
\newcommand{\etaKst}{\ensuremath{\B\ra\fetaKst}}

\newcommand{\etaKstp}{\ensuremath{\Bp\ra\fetaKstp}}
\newcommand{\BetaKstp}{\ensuremath{\calB(\etaKstp)}}
\newcommand{\retaKstp}{\ensuremath{xx^{+xx}_{-xx}\pm xx}}

\newcommand{\AetaKstp}{\ensuremath{xx\pm xx\pm xx}}
\newcommand{\setaKstp}{\ensuremath{xx}}

\newcommand{\etaKstz}{\ensuremath{\Bz\ra\fetaKstz}}
\newcommand{\BetaKstz}{\ensuremath{\calB(\etaKstz)}}
\newcommand{\retaKstz}{\ensuremath{xx^{+xx}_{-xx}\pm xx}}

\newcommand{\AetaKstz}{\ensuremath{xx\pm xx \pm xx}}
\newcommand{\setaKstz}{\ensuremath{xx}}

  \newcommand{\fetapppKstz}{\ensuremath{\eta_{3\pi} K^{*0}}}

  \newcommand{\KstOne}{\ensuremath{K^{*}(892)}}

  \newcommand{\etaKststz}{\ensuremath{\etaK_0^{*0}}(\textit{S}-\textit{wave})}
  \newcommand{\etaKststp}{\ensuremath{\etaK_0^{*+}}(\textit{S}-\textit{wave})}

  \newcommand{\fetaggKTstpKspip}{\ensuremath{\eta_{\gamma\gamma}K^{*+}_0(\KS \pi^+)}}
  \newcommand{\fetaggKTstpKppiz}{\ensuremath{\eta_{\gamma\gamma}K^{*+}_0(\Kp \pi^0)}}
  \newcommand{\fetaggKTstzKppim}{\ensuremath{\eta_{\gamma\gamma}K^{*0}_0(\Kp \pi^-)}}
  \newcommand{\fetapppKTstpKspip}{\ensuremath{\eta_{3\pi}K^{*+}_0(\KS \pi^+)}}
  \newcommand{\fetapppKTstpKppiz}{\ensuremath{\eta_{3\pi}K^{*+}_0(\Kp \pi^0)}}
  \newcommand{\fetapppKTstzKppim}{\ensuremath{\eta_{3\pi}K^{*0}_0(\Kp \pi^-)}}
  \newcommand{\fetaggKstpKspipP}{\ensuremath{\eta_{\gamma\gamma}K^{*+}_{\KS\pi^+}}(892)}
  \newcommand{\fetaggKstpKppizP}{\ensuremath{\eta_{\gamma\gamma}K^{*+}_{\Kp \pi^0}}(892)}
  \newcommand{\fetaggKstzKppimP}{\ensuremath{\eta_{\gamma\gamma}K^{*0}_{\Kp \pi^-}}(892)}
  \newcommand{\fetapppKstpKspipP}{\ensuremath{\eta_{3\pi}K^{*+}_{\KS \pi^+}}(892)}
  \newcommand{\fetapppKstpKppizP}{\ensuremath{\eta_{3\pi}K^{*+}_{\Kp \pi^0}}(892)}
  \newcommand{\fetapppKstzKppimP}{\ensuremath{\eta_{3\pi}K^{*0}_{\Kp \pi^-}}(892)}
  \newcommand{\fetaggKstpKspipD}{\ensuremath{\eta_{\gamma\gamma}K^{*+}_2(\KS \pi^+)}}
  \newcommand{\fetaggKstpKppizD}{\ensuremath{\eta_{\gamma\gamma}K^{*+}_2(\Kp \pi^0)}}
  \newcommand{\fetaggKstzKppimD}{\ensuremath{\eta_{\gamma\gamma}K^{*0}_2(\Kp \pi^-)}}
  \newcommand{\fetapppKstpKspipD}{\ensuremath{\eta_{3\pi}K^{*+}_2(\KS \pi^+)}}
  \newcommand{\fetapppKstpKppizD}{\ensuremath{\eta_{3\pi}K^{*+}_2(\Kp \pi^0)}}
  \newcommand{\fetapppKstzKppimD}{\ensuremath{\eta_{3\pi}K^{*0}_2(\Kp \pi^-)}}

\newcommand{\BABARPubYear}    {06}
\newcommand{\BABARPubNumber}  {053}

\newcommand{\SLACPubNumber} {12011}
\newcommand{\LANLNumber} {0608005}

\newcommand{\KstZero}{\ensuremath{K_0^*(1430)}}
\newcommand{\KstTwo}{\ensuremath{K_2^*(1430)}}

\renewcommand{\etaKstz}{\ensuremath{\Bz\ra\eta K^{*0}(892)}}
\newcommand{\KpiSwave}{\ensuremath{(K\pi)^*_0}}
\newcommand{\KpizSwave}{\ensuremath{(K\pi)^{*0}_0}}
\newcommand{\KpipSwave}{\ensuremath{(K\pi)^{*+}_0}}

\renewcommand{\etaKststz}{\ensuremath{\Bz\ra\eta\KpizSwave}}

\newcommand{\etaKstTwoz}{\ensuremath{\Bz\ra\eta K_2^{*0}}(1430)}

\renewcommand{\etaKstp}{\ensuremath{\Bp\ra\eta K^{*+}(892)}}

\renewcommand{\etaKststp}{\ensuremath{\Bp\ra\eta\KpipSwave}}

\newcommand{\etaKstTwop}{\ensuremath{\Bp\ra\eta K_2^{*+}}(1430)}

\renewcommand{\retaKstz}{\ensuremath{16.5\pm1.1\pm0.8}}
\renewcommand{\AetaKstz}{\ensuremath{0.21\pm0.06\pm 0.02}}
\renewcommand{\setaKstz}{\ensuremath{18.8}}
\renewcommand{\BetaKstz}{\ensuremath{{\cal B}(B^0\ra\eta K^{*0}(892))}}
\renewcommand{\retaKstp}{\ensuremath{18.9\pm1.8\pm1.3}}
\renewcommand{\AetaKstp}{\ensuremath{0.01\pm 0.08\pm 0.02}}
\renewcommand{\setaKstp}{\ensuremath{13.0}}
\renewcommand{\BetaKstp}{\ensuremath{{\cal B}(B^+\ra\eta K^{*+}(892))}}
 
\newcommand{\retaKststz}{\ensuremath{11.0\pm1.6\pm 1.5}}
\newcommand{\AetaKststz}{\ensuremath{0.06\pm 0.13\pm 0.02}}
\newcommand{\setaKststz}{\ensuremath{5.7}}
\newcommand{\BetaKststz}{\ensuremath{{\cal B}}(\textrm{\etaKststz})}
\newcommand{\retaKststp}{\ensuremath{18.2\pm2.6\pm 2.6}}
\newcommand{\AetaKststp}{\ensuremath{0.05\pm 0.13\pm 0.02}}
\newcommand{\setaKststp}{\ensuremath{5.9}}
\newcommand{\BetaKststp}{\ensuremath{{\cal B}}(\textrm{\etaKststp})}
 
\newcommand{\retaKstTwoz}{\ensuremath{9.6\pm1.8\pm 1.1}}
\newcommand{\AetaKstTwoz}{\ensuremath{-0.07\pm0.19\pm 0.02}}
\newcommand{\setaKstTwoz}{\ensuremath{5.3}}
\newcommand{\BetaKstTwoz}{\ensuremath{{\cal B}}(\textrm{\etaKstTwoz})}
\newcommand{\retaKstTwop}{\ensuremath{9.1\pm2.7\pm 1.4}}
\newcommand{\AetaKstTwop}{\ensuremath{-0.45\pm 0.30\pm 0.02}}
\newcommand{\setaKstTwop}{\ensuremath{3.5}}
\newcommand{\BetaKstTwop}{\ensuremath{{\cal B}}(\textrm{\etaKstTwop})}

\begin{document}

\preprint{\babar-PUB-\BABARPubYear/\BABARPubNumber} 
\preprint{SLAC-PUB-\SLACPubNumber} 


\begin{flushleft}
\babar-PUB-\BABARPubYear/\BABARPubNumber\\
SLAC-PUB-\SLACPubNumber\\
hep-ex/\LANLNumber\\[10mm]
\end{flushleft}

\title{
\large \bf \boldmath
Measurement of branching fractions and charge asymmetries in $B$ decays to
an $\eta$ meson and a \Kstar\ meson }

%
\author{B.~Aubert}
\author{M.~Bona}
\author{D.~Boutigny}
\author{F.~Couderc}
\author{Y.~Karyotakis}
\author{J.~P.~Lees}
\author{V.~Poireau}
\author{V.~Tisserand}
\author{A.~Zghiche}
\affiliation{Laboratoire de Physique des Particules, IN2P3/CNRS et Universit\'e de Savoie,
 F-74941 Annecy-Le-Vieux, France }
\author{E.~Grauges}
\affiliation{Universitat de Barcelona, Facultat de Fisica, Departament ECM, E-08028 Barcelona, Spain }
% ...
\collaboration{The \babar\ Collaboration}
\noaffiliation

\date{\today}

\begin{abstract}
We present measurements of branching fractions and charge asymmetries for the decays
\etaKst, where \Kstar\ indicates a spin 0, 1, or 2 $K\pi$ system.
The data sample corresponds
to 344$\times10^6$ \BB\ pairs collected with the \babar\ detector 
at the PEP-II asymmetric-energy \epem\ collider at SLAC.
We measure the branching fractions (in units of $10^{-6}$):
$\BetaKstz=\retaKstz$,~
$\BetaKstp=\retaKstp$,~
$\BetaKststz=\retaKststz$,
$\BetaKststp=\retaKststp$,
$\BetaKstTwoz=\retaKstTwoz$,
and~
$\BetaKstTwop=\retaKstTwop$.
We also determine the charge asymmetries for all decay modes.
\end{abstract}

\pacs{13.25.Hw, 11.30.Er}

\maketitle


Decays of $B$ mesons to charmless hadronic final states are widely used to 
test the accuracy of theoretical predictions.  The decays involving 
$\eta$ and \etapr\ mesons have received considerable attention since early
predictions were unable to explain the data.  For decays of interest in 
this paper, there have been recent calculations from QCD factorization
\cite{BNgeneral,etapQCDfact} and flavor~SU(3) symmetry~\cite{chiangGlob}.

Charmless $B$ decays to final states with strangeness are expected to be dominated by
$b\to s$ loop (``penguin") amplitudes.  The branching fraction for the decay
\etaKst\ is expected to be larger than most similar decays (though not as 
large as $\B\ra\etapr K$)
due to constructive interference between two penguin amplitudes~\cite{Lipkin}.

While the decay \etaKst(892) has been seen previously \cite{CLEO,PRD}, there 
have been no searches for states with an $\eta$ meson accompanied
by $K^{*}(1430)$ mesons, and no theoretical predictions
exist for these decays.  In this Letter we present 
measurements of branching fractions and charge asymmetries for the decays
\etaKstz\ \cite{conjugate}, \etaKstp, \etaKststz, \etaKststp, \etaKstTwoz, 
and \etaKstTwop,
where we denote by \KpiSwave\ the $0^+$ component of the $K\pi$ spectrum.
The charge asymmetry is defined as
$\acp \equiv (\Gamma^--\Gamma^+)/(\Gamma^-+\Gamma^+)$, where the superscript 
on the width $\Gamma$ corresponds to the sign of the $B^\pm$ 
meson or the sign of the charged kaon for \Bz\ decays.


The results presented here are obtained from 
data collected with the \babar\ detector (described 
in detail elsewhere~\cite{BABARNIM}) at the PEP-II asymmetric \epem
collider located at the Stanford Linear Accelerator Center.
The analysis uses an integrated luminosity of 313~fb$^{-1}$, corresponding
to 344$\times10^6$ \BB\ pairs, recorded
at the $\Upsilon (4S)$ resonance (center-of-mass (CM) energy $\sqrt{s}=10.58\
\gev$), and follows closely the technique described in detail
in Ref.~\cite{PRD}.
The sample is 3.9 times larger than that of Ref.~\cite{PRD}.


The \Kstar\ mesons are reconstructed from $K^+\piz$ (\KstpKppiz),
$\KS\pip$ (\KstpKspip), or $K^+\pim$ (\KstzKppim) final states.
All tracks from resonance candidates are required to have charged
particle identification (PID) consistent with kaons or pions.
We select $\eta$, \kzs\ and $\piz$ candidates from the decays \etatogg\ 
(\etagg), \etatoppp\ (\etappp), $\kzs\ra\pip\pim$ and $\piz\ra\gaga$.  We
impose the following requirements on the invariant masses (in \mev) of the 
particle candidate final states: $490< m_{\gaga}<600$ for 
\etagg, $520<m_{\pi\pi\pi}< 570$ for \etappp, $486< m_{\pi\pi}<510$ for \kzs\
and $120 < m_{\gaga} < 150$ for \piz.  For \kzs\ candidates we require at least
3 standard-deviation ($\sigma$) three-dimensional separation
between the decay vertex and the \epem\ collision point.  
Requirements are loose 
for the variables used in the maximum likelihood (ML) fit described below.
For the $K\pi$ system, we define a low-mass region (LMR) by $755<m_{K\pi}<1035$ \mev\
and a high-mass region (HMR) by $1035<m_{K\pi}<1535$ \mev \cite{lohimass}.
For the \Kstar\ we use the helicity frame,
defined as the \Kstar\ rest frame with polar axis opposite to the direction of 
the $B$.  We define $\hel\equiv\cos{\theta_H}$,
where the decay angle $\theta_H$ is the polar angle of the kaon momentum in 
the helicity frame.  For the LMR, we require $-0.95<\hel<1.0$ for
\Kstarz\ and \KstpKspip, and $-0.7<\hel<1.0$ for \KstpKppiz.  For the
HMR, we require $-0.5<\hel<1.0$ for all modes in order to remove the
region in \hel\ having very large backgrounds.

A $B$-meson candidate is characterized kinematically by the
energy-substituted mass
$\mes=(\frac{1}{4}s-\pvec_B^2)^\half$
and energy difference $\DE = E_B-\half\sqrt{s}$, where
$(E_B,\pvec_B)$ is the $B$-meson 4-momentum vector, and
all values are expressed in the \UfourS\ frame.
Signal events peak at zero for \DE, and at the $B$ mass \cite{PDG2004} 
for \mes, with a resolution for \DE\ (\mes) of 30-45 MeV ($3.0\ \mev$).
We require $|\DE|\le0.2$ GeV and $5.25\le\mes<5.29\ \gev$.

The angle \thetaT\ between the thrust axis of the $B$ candidate in the \UfourS\
frame and that of the rest of the charged tracks and neutral clusters in the 
event is used to reject the dominant continuum $\epem\ra\qqbar$
($q=u,d,s,c$) background events.  The distribution of $|\costhr|$ is
sharply peaked near $1.0$ for combinations drawn from jet-like \qqbar\
pairs, and nearly uniform for the almost isotropic $B$-meson decays;
we require $|\costhr|\le0.9$. Further discrimination
from continuum in the ML\ fit is obtained from energy flow in the
event via a Fisher discriminant \xf\
that is described in detail elsewhere~\cite{PRD}.

For the modes with \etatogg, we reject $B\ra\Kstar\gamma$ background with 
the requirement $|\cos\theta^\eta_{\rm dec}|\le0.86$, where $\theta^\eta_{\rm dec}$
is the $\eta$ decay angle defined, in the $\eta$ rest frame, as the angle between
one of the photons and the $B$ direction.

When there are multiple candidates (less than 30\% of events \cite{lohimass}),
we choose the candidate with a value of the reconstructed $\eta$ mass closest 
to the PDG mass \cite{PDG2004}.


We use Monte Carlo (MC) simulations~\cite{geant} for
the few charmless \BB\ background decays that
survive the candidate selection and have characteristics similar to the signal.
We find these contributions to be negligible for all modes with an \etatoppp\ 
decay except \fetapppKstz.  For all other modes, we include a 
component in the ML fit to account for them.


We obtain yields and \acp\ for each decay chain from an extended unbinned 
maximum likelihood 
fit with the following input observables: \DE, \mes, \xf,  
$\mres$ (the masses of the $\eta$ and \Kstar\ candidates), and \hel.
For each event $i$ and hypothesis $j$ (signal, continuum background, 
\BB\ background), we define the  probability density function (PDF),
with resulting likelihood \calL:
\begin{eqnarray}
{\cal P}^i_{j}&=& {\cal P}_j (\mes^i) {\cal  P}_j (\DE^i) 
 { \cal P}_j(\xf^i) {\cal P}_j (\mres^i) {\cal P}_j (\hel^i)\\
{\cal L}&=&\exp{(-\sum_{j} Y_{j})}
\prod_i^{N}\left[\sum_{j} Y_{j} {\cal P}^i_{j}\right]\,,
\end{eqnarray}
where $Y_{j}$ is the yield of events of hypothesis $j$, and $N$ is the
number of events in the sample. 
The free parameters of the fit are the signal and background yields,
between 9 and 11
\qqbar\ background PDF parameters (see below), and the signal and \qqbar
background charge asymmetries.

We determine the contributions from \KstOne, \KpiSwave, and \KstTwo\ by fits
in the LMR and HMR.  The fit in the LMR
includes \KstOne\ and \KpiSwave\ signal components (\KstTwo\ is negligible
in this region), with the fixed \KpiSwave\ yield determined from the result 
of the fit to the HMR.  For the fit in the HMR, all three 
components are included; the \KstOne\ yield is fixed from the result of the
fit in the LMR, while the \KpiSwave\ and \KstTwo\ branching 
fractions are free in a simultaneous fit over the two (four) sub-decay modes
for \Kstarz\ (\Kstarp).  For the generated \KpiSwave\ spectrum,
we use the LASS parameterization~\cite{LASS} which consists of the
\KstZero\ resonance together with an effective-range non-resonant component.
The \KstTwo\ is generated as a relativistic Breit-Wigner shape with known mass 
and width \cite{PDG2004}.


For the signal and \BB\ background components we determine the PDF
parameters from MC.
For background from continuum and non-peaking combinations from $B$ decays,
we obtain the PDF from (\mb,\,\DE) sideband data for each decay,
before applying the fit
to data in the signal region; we refine this PDF by letting all parameters
vary in the final fit.  
We parameterize each of the functions ${\cal P}_{\rm
sig}(\mes),\ {\cal P}_{\rm sig}(\DE),\ { \cal P}_j(\xf)$
and the peaking components of ${\cal P}_j(\mres)$ with either
a Gaussian, the sum of two Gaussians or an asymmetric Gaussian function
as required to describe the distribution.  For ${\cal P}_{\rm sig}(\hel)$
we use a low order polynomial.  Slowly varying distributions
(all masses, \DE\ and \hel\ for continuum background) are
represented by one or a combination of linear, quadratic and phase-space
motivated functions~\cite{PRD}.  
The fitted \qqbar\ background PDF parameters are
found to be in close agreement with the initial values.
Control samples with topologies similar to
our signal modes (e.g.\ $B\ra D(K\pi\pi)\pi$) are used
to calibrate the simulated resolutions evaluated from~MC~\cite{PRD}.


\begin{table*}[btp]
\caption{
Fitted signal yield $Y_S$ in events (ev.), measured bias (see text), detection
efficiency $\epsilon$, daughter branching fraction product ($\prod\calB_i$),
significance~\calS\ (with systematic uncertainties included), measured 
branching fraction \calB, and signal charge asymmetry \acp\ for each mode.
The first uncertainty is statistical and  the second systematic.
}
\label{tab:results}
\begin{tabular}{lcccccccc}
\dbline
Mode	      	& $Y_S$	(ev.) & Bias (ev.) &$\epsilon$ (\%) &$\prod\calB_i$ (\%)
		& $\cal S$ ($\sigma$)	&  \calB\ $(10^{-6})$	& \acp\	\\
\tbline
~~\fetaggKstzKppimP	&$407\pm29$ & $+15$ &24	&26  &17.6  &$18.2\pm1.4$&$\msp0.24\pm0.07$ \\
~~\fetapppKstzKppimP	&$111\pm16$ & $+13$ &16	&15  &~6.3  &$10.9\pm2.0$&$\msp0.12\pm0.14$ \\
\bma{\etaKstz}	&       &                  &  	   &  	&\bma{\setaKstz}&\bma{\retaKstz}&\bma{\msp\AetaKstz}  \\
~~\fetaggKstpKppizP	&$~99\pm16$ & $+~7$ &11	&13  &~6.9 &$18.0\pm3.2$&$\msp0.19\pm0.16$ \\
~~\fetapppKstpKppizP	&$~56\pm11$ & $+~4$ &~8	&~8  &~6.1 &$25.4\pm5.5$&$-0.05\pm0.20$ \\
~~\fetaggKstpKspipP	&$149\pm19$ & $+12$ &22	&~9  &~8.6 &$20.5\pm2.9$&$-0.03\pm0.13$ \\
~~\fetapppKstpKspipP	&$~36\pm10$ & $+~5$ &15	&~5  &~3.8 &$11.9\pm3.9$&$-0.23\pm0.28$ \\
\bma{\etaKstp}	&       &  		   &  	   &	&\bma{\setaKstp} &\bma{\retaKstp}&\bma{\msp\AetaKstp}    \\
~~\fetaggKTstzKppim	&$163\pm25$ & $+17$ &15	&26  &~5.3	&$10.8\pm1.9$ &$\msp0.14\pm0.15$ \\
~~\fetapppKTstzKppim	&$~69\pm17$ & $+~9$ &10	&15  &~3.6	&$11.4\pm3.2$ &$-0.18\pm0.25$ \\
\bma{\etaKststz}	&&    &  &  	&\bma{\setaKststz} &\bma{\retaKststz}	& \bma{\msp\AetaKststz} \\
~~\fetaggKTstpKppiz	&$~93\pm20$& $+~9$  &10	&13  &~4.3	&$19.2\pm4.5$ &$-0.05\pm0.21$ \\
~~\fetapppKTstpKppiz	&$~39\pm12$& $+~6$  &~7	&~8  &~3.4	&$18.0\pm6.3$ &$\msp0.03\pm0.29$ \\
~~\fetaggKTstpKspip	&$~55\pm16$& $+~5$  &12	&~9  &~3.0	&$13.3\pm4.2$ &$\msp0.13\pm0.25$ \\
~~\fetapppKTstpKspip	&$~49\pm11$& $+~3$  &~9	&~5  &~4.4	&$28.1\pm6.7$ &$\msp0.18\pm0.22$ \\
\bma{\etaKststp}&	& &  	&  	&\bma{\setaKststp} 	&\bma{\retaKststp}& \bma{\msp\AetaKststp}  \\
~~\fetaggKstzKppimD	&$~72\pm17$& $-~1$  &18 & 14 &~4.7 &$~8.4\pm1.9$ &$-0.20\pm0.23$& \\
~~\fetapppKstzKppimD	&$~40\pm13$& $-~1$  &12 & ~8 &~3.4 &$12.5\pm4.1$ &$\msp0.23\pm0.31$ \\
\bma{\etaKstTwoz}	&  & & &  &\bma{\setaKstTwoz}	&\bma{~\retaKstTwoz}& \bma{\AetaKstTwoz}  \\
~~\fetaggKstpKppizD	&$~26\pm12$& $-~1$  &13 & ~7 &~2.3 &$~9.1\pm4.0$ &$-0.16\pm0.41$ \\
~~\fetapppKstpKppizD	&$~20\pm~8$& $-~1$  &~9 & ~4 &~2.6 &$17.8\pm7.2$ &$-0.82\pm0.47$ \\
~~\fetaggKstpKspipD	&$~12\pm10$& $-~1$  &13 & ~5 &~1.8 &$~6.4\pm4.7$ &$\msp0.05\pm0.58$ \\
~~\fetapppKstpKspipD	&$~~2\pm~5$& $+~1$  &10 & ~3 &~0.2 &$~0.9\pm5.1$ &$-1.00\pm1.56$\\
\bma{\etaKstTwop}	& & &  	&  	&\bma{\setaKstTwop}  &\bma{~\retaKstTwop}& \bma{\AetaKstTwop}\\
\dbline
\end{tabular}
\end{table*}

Before applying the fitting procedure to the data we subject it to
several tests.
In particular, we evaluate possible biases in the yields from our neglect of
small residual correlations among 
discriminating variables in the signal and charmless \BB\ background PDFs.
The bias is determined by fitting ensembles of simulated \qqbar\ experiments 
generated from the PDFs into which we have embedded the expected number of 
signal and \BB\ background events, randomly extracted from the fully simulated 
MC samples.  The small biases are listed in Table~\ref{tab:results}. 
We measure the correlations in the data and find them to be negligibly small. 


We compute the branching fraction for each decay by
subtracting the fit bias from the measured yield, and dividing the
result by the efficiency and the number of produced \BB\ pairs.
We assume equal decay rates for the \UfourS\ to \BpBm\ and \BzBzb .
In Table~\ref{tab:results} we show for each decay mode the measured
branching fraction together with the event yield $Y_S$, efficiency $\epsilon$, 
and \acp.  The significance is taken as the square root of the difference 
between the value of $-2\ln{\cal L}$
(with systematic uncertainties included)
for zero signal and the value at its minimum.


For the LMR
the measurements for separate daughter decays are
combined by adding the values of $-2\ln{\cal L}$ as functions of
the branching fractions, taking account of the correlated and 
uncorrelated systematic uncertainties~\cite{PRD} described below.

\begin{figure}[!htb]
 \includegraphics[angle=0,scale=0.4]{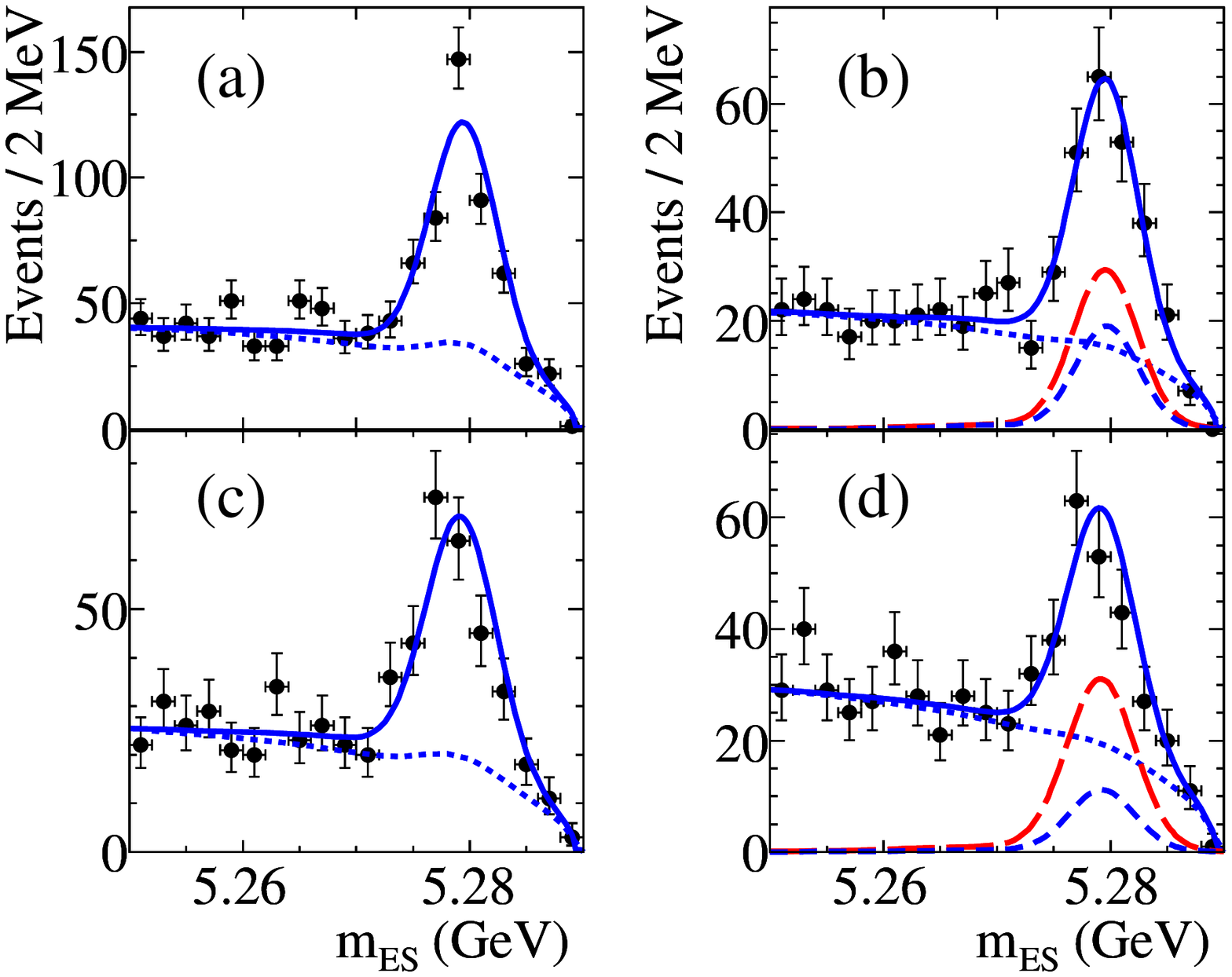}
\vspace{-0.2cm}
 \caption{\label{fig:projMES}
$B$-candidate \mb\ projections obtained with
a cut on the signal likelihood (see text) for (a) \etaKstz,
(b) \etaKststz\ (long-dashed, red) plus \etaKstTwoz\ (short-dashed, blue), 
(c) \etaKstp,
and (d) \etaKststp\ (long-dashed, red) plus \etaKstTwoz\ (short-dashed, blue).
Points with uncertainties represent the data, solid curves the full fit 
functions, and dotted curves the full background functions.}
\end{figure}

\begin{figure}[!htb]
 \includegraphics[angle=0,scale=0.4]{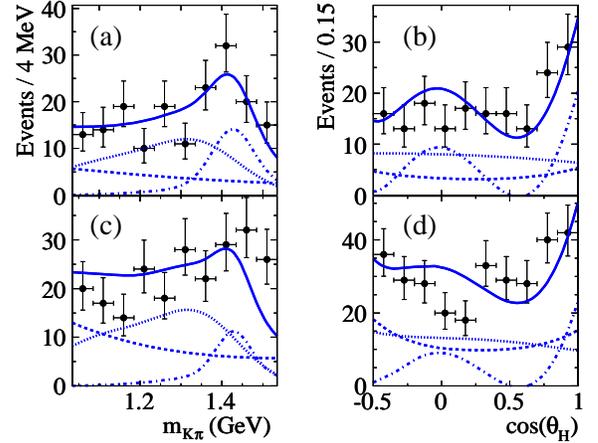}
\vspace{-0.2cm}
 \caption{\label{fig:projHMR}
Projection of the signals in the HMR, obtained with
a cut on the signal likelihood (see text): $K\pi$ mass for (a) \Bz, and (c)
\Bp channels; \hel\ for (b) \Bz, and (d) \Bp channels.  Points with
uncertainties represent the data, solid curves the full fit functions,
dotted curves the \KpiSwave\ portion, dot-dashed curves the \KstTwo\ portion, 
and dashed curves the full background functions.}
\end{figure}

In Fig.~\ref{fig:projMES} we show projections onto \mes\
of subsamples enriched with a threshold requirement on
the signal likelihood (computed without the variable plotted) that optimizes 
the sensitivity.  
There are substantial signals in all four samples.  For the
HMR, separation of the \KpiSwave\ and \KstTwo\ signals is
afforded mainly by the $K\pi$ mass and helicity shapes; projections of
these distributions are shown in Fig.~\ref{fig:projHMR}. The
statistical correlations between the two signals are $\sim$$-0.42$ in the HMR 
fits to both the \Bz\ and \Bp\ decays.

The largest systematic uncertainties are due to the signal and \BB\
PDF modeling, the fit bias correction, the modeling of the $K\pi$ mass
distribution, the neutral selection efficiency, and neglect of
interference between signal components.
The PDF modeling error is largely included in the statistical uncertainty
since all background parameters are free in the fit.  The uncertainties in
the signal PDF parameters are estimated from the consistency of fits to MC
and data in control samples with similar final states.  Varying the signal
PDF parameters  within these errors, we estimate the mode-dependent
uncertainties to be 1--4 events.  The uncertainty in the fit bias correction 
is taken to be half of the correction.
We estimate the uncertainty from modeling the \BB\ backgrounds 
to be less than 1 event.

Uncertainties in the reconstruction efficiency, found from auxiliary 
studies of inclusive control samples~\cite{PRD}, are $0.4\%$ per track,
$3.0\%$ per $\eta/\pi^0$, and 1.9\%\ for a \KS.  Our estimate of 
the systematic uncertainty for the number of \BB\ pairs is 1.1\%.
Published data~\cite{PDG2004}\ provide the
uncertainties for the $B$-daughter product branching fractions (1--2\%).
The uncertainty due to the efficiency of the \costhr\ requirement is 0.5\%.
The systematic uncertainty for \acp\ is estimated to be 2\%, dominated
by tracking and PID systematic effects \cite{phiKstarPRL}.

Since our model does not account
for interference among the components, we assign systematic
uncertainties based on the $m(K\pi)$-dependence of the complex phases
measured in Ref.~\cite{LASS}, with allowance for unknown process-dependent
overall phases.  The effect is small for the LMR and about 10\%
for the HMR.  For the HMR, the systematic uncertainties are applied
after the combined fit, taking sub-mode errors as correlated.


In summary, we have presented improved measurements of the branching 
fractions for the decays \etaKstz\ and \etaKstp, as well as measurements
of the decays \etaKststz, \etaKststp, \etaKstTwoz, and \etaKstTwop,
which had not been seen previously.  The first two supersede previous
\babar\ measurements \cite{PRD} and agree with earlier results 
and theoretical predictions \cite{BNgeneral,etapQCDfact,chiangGlob}.  
We also calculate the branching fraction for the resonant decays to
$\eta\KstZero$ using the composition of \KpiSwave\ from \cite{LASS}.  We
find $\calB(\Bz\to\eta K_0^{*0}(1430))=(7.8\pm1.1\pm0.6\pm0.9)\timesix$
and $\calB(\Bp\ra\eta K_0^{*+}(1430))=(12.9\pm1.8\pm1.1\pm1.4)\timesix$,
where the third errors arise from the uncertainties on the branching 
fraction $K^*_0(1430)\ra K\pi$ \cite{PDG2004} and the resonant fraction of 
\KpiSwave.

There are no theoretical predictions for the decays involving spin-0 or 2 mesons.
The measured values of \acp\ are mostly consistent with zero within their 
uncertainties; the value for \etaKstz\ shows evidence for direct \CP\ violation.


We are grateful for the excellent luminosity and machine conditions
provided by our \pep2\ colleagues, 
and for the substantial dedicated effort from
the computing organizations that support \babar.
The collaborating institutions wish to thank 
SLAC for its support and kind hospitality. 
This work is supported by
DOE
and NSF (USA),
NSERC (Canada),
IHEP (China),
CEA and
CNRS-IN2P3
(France),
BMBF and DFG
(Germany),
INFN (Italy),
FOM (The Netherlands),
NFR (Norway),
MIST (Russia),
MEC (Spain), and
PPARC (United Kingdom). 
Individuals have received support from the
Marie Curie EIF (European Union) and
the A.~P.~Sloan Foundation.


\begin{thebibliography}{99}
\bibitem{conjugate}
Except as noted explicitly, we use a particle name to denote either member
of a charge conjugate pair.
\end{thebibliography}
\end{document}